\documentstyle[aaspp4,emulateapj5,apjfonts]{article}

\gdef\h50min{$h_{50}^{-1}$}
\gdef\h65{$h_{65}^{-1}$}
\gdef\kms{km\,s$^{-1}$}
\gdef\hk{Ca\,{\sc ii} H and K}
\gdef\mgb{Mg$_b$}
\gdef\3727{[O\,{\sc ii}]\,3727\,\AA}
\gdef\5007{[O\,{\sc iii}]\,5007\,\AA}
\gdef\oii{[O\,{\sc ii}]}
\gdef\clusa{RDCS~J0848+4453}
\gdef\clusb{RDCS~J0848+4456}
\gdef\clusc{RDCS~J0848+4452}
\gdef\gala{J0848-1}
\gdef\galb{J0848-2}
\gdef\galc{J0848-3}
\lefthead{van Dokkum \& Stanford}
\righthead{Fundamental Plane at $z=1.27$}
\slugcomment{Accepted for publication in the Astrophysical Journal}
\begin{document}

\title{The Fundamental Plane at $z=1.27$: First Calibration of the
Mass Scale of Red Galaxies at Redshifts $z>1$
\altaffilmark{1,2}}
\author{Pieter G. van Dokkum}
\affil{California Institute of Technology, MS105-24, Pasadena,
CA 91125}
\author{\and S. A. Stanford\altaffilmark{3}}
\affil{Physics Department, University of California-Davis, Davis, CA 
95616}
\altaffiltext{1}{Based on observations with the NASA/ESA {\em Hubble Space
Telescope}, obtained at the Space Telescope Science Institute, which
is operated by AURA, Inc., under NASA contract NAS 5--26555.}
\altaffiltext{2}
{Based on observations obtained at the W.\ M.\ Keck Observatory,
which is operated jointly by the California Institute of
Technology and the University of California.}
\altaffiltext{3}{Institute of Geophysics and Planetary Physics, Lawrence
Livermore National Laboratory}

\begin{abstract}

We present results on the Fundamental Plane of early-type galaxies
in the cluster \clusa\ at $z=1.27$.
Internal velocity dispersions of three $K$-selected early-type galaxies
are determined from deep Keck spectra,
using absorption lines in the rest-frame wavelength
range 3400\,\AA\ -- 4000\,\AA. Structural parameters
are determined from {\em Hubble Space Telescope} NICMOS images.
The galaxies show substantial offsets from the Fundamental
Plane of the nearby Coma cluster,
as expected from passive evolution of their stellar populations.
The offsets from the Fundamental Plane can be expressed as
offsets in mass-to-light ($M/L$) ratio.
The
$M/L$ ratios of the two most massive galaxies
are consistent with
an extrapolation of results obtained for clusters at $0.02<z<0.83$.
The evolution of early-type galaxies with masses $>10^{11}\,M_{\odot}$
is well described by $\ln M/L_B \propto (-1.06\pm 0.09) z$,
corresponding to passive evolution of $-1.50 \pm 0.13$ magnitudes
at $z=1.3$. Ignoring selection effects,
the best fitting stellar formation redshift is
$z_*=2.6^{+0.9}_{-0.4}$ for an $\Omega_m=0.3$, $\Omega_{\Lambda}=0.7$ cosmology
and a Salpeter IMF,
corresponding to a luminosity weighted age at the
epoch of observation of $\sim 2$\,Gyr.
The $M/L$ ratios of these two galaxies
are also in excellent agreement with predictions from
models that include selection effects caused by morphological
evolution (``progenitor bias'').
The third galaxy is a factor $\sim 10$ less
massive than the other two,
shows strong Balmer absorption lines in its spectrum,
and is offset from the Coma Fundamental Plane by $2.9$
mag in rest-frame $B$. Larger samples are required to determine
whether such young early-type galaxies are common in high redshift
clusters. Despite their large range in $M/L$ ratios, all three
galaxies fall in the ``Extremely Red Object'' (ERO) class with $I-H>3$ and
$R-K>5$, and our results show that it is hazardous to use simple models
for converting luminosity to mass for these objects.
The work presented here, and previous mass measurements at lower
redshift, can be considered first steps to
empirically disentangle
luminosity and mass evolution at the high mass end of the
galaxy population, lifting an important degeneracy in the
interpretation
of evolution of the luminosity function.

\end{abstract}

\keywords{
galaxies: clusters: general ---
galaxies: evolution ---
galaxies: structure of ---
galaxies: kinematics and dynamics ---
galaxies: elliptical and lenticular, cD
}

\section{Introduction}

One of the defining characteristics of galaxy formation models is
the evolution of the galaxy mass function with time.
In traditional ``monolithic collapse'' models, galaxies formed
at early times from the collapse of proto-galactic gas clouds, and
the mass function evolved very little after $z\sim 3$
~(e.g., {Eggen}, {Lynden-Bell}, \&  {Sandage} 1962; {Searle}, {Sargent}, \&  {Bagnuolo} 1973; {Jimenez} {et~al.} 1999).
In contrast, hierarchical models for galaxy
formation in Cold Dark Matter cosmologies postulate that
galaxies continuously form in mergers,
implying strong evolution in the mass function over the redshift range
$0<z<2$ ~(e.g., {White} \& {Frenk} 1991; {Kauffmann} \& {Charlot} 1998; {Cole} {et~al.} 2000; {Somerville}, {Primack}, \&  {Faber} 2001).

Although these models are very different, determining which
is closer to reality has proven to be quite difficult.
An important constraint on the evolution of the mass function is
the evolution of the
$K$-band luminosity function, under the assumption that
$K$-band light is a good tracer of stellar mass to $z \gtrsim 1$
~({Kauffmann} \& {Charlot} 1998). 
Observationally, this evolution is difficult to measure as deep
$K$-selected redshift surveys over large fields are required.
Past and ongoing
surveys seem to indicate some evolution in the number density,
although the amount is still uncertain
~(e.g., {Cowie} {et~al.} 1996; {Kauffmann} \& {Charlot} 1998; {Stern} {et~al.} 2001; Drory et~al.\ 2001; {Cimatti} {et~al.} 2002).
Other studies focus exclusively on the reddest galaxies, usually defined
as having $I-H>3$ or $R-K>5$ ~(e.g., {Daddi}, {Cimatti}, \& {Renzini} 2000; {McCarthy} {et~al.} 2001; {Roche} {et~al.} 2002).
A large fraction of these ``Extremely Red Objects'' (EROs)
are thought to be passively evolving
massive galaxies at $1\lesssim z \lesssim 2$, whose number density may
constrain the evolution of the mass function without the need for full
redshift information.

One of the main uncertainties in the interpretation is the
model-dependent conversion of luminosity to stellar mass.  The
observed evolution of the luminosity function is caused by a
combination of evolution of the underlying mass function and the
evolution of stellar populations. At $z \sim 1$, the $K$-band samples
the rest-frame $J$-band, and mass-to-light ($M/L$) ratios of galaxies
are quite uncertain even if they are ``pre-selected'' to be passively
evolving through the use of broad-band colors.  As an example, the
~{Worthey} (1994)
models\footnote{www.astro.wsu.edu/worthey/dial/dial\_a\_model.html}
predict a variation of a factor $2.4$ in the $M/L_J$ ratio for
luminosity weighted ages ranging from 1 -- 5 Gyr and [Fe/H]
ranging from $-0.225$ to $+0.5$. At the bright end of
the luminosity function this uncertainty in $M/L$ ratio contributes
directly to the uncertainty in the number density.

For a correct interpretation of the luminosity function it is
therefore essential to measure masses and mass-to-light ratios
directly, so that luminosities can be converted to masses in a
model-independent way.  Total masses of galaxies are notoriously
difficult to measure, but scaling relations such as the Tully-Fisher
relation for spiral galaxies ~({Tully} \& {Fisher} 1977) and the
Fundamental Plane (FP) for early-type galaxies
~({Djorgovski} \& {Davis} 1987; {Dressler} {et~al.} 1987) can be used to constrain the
evolution of the $M/L$ ratio relative to local samples ~(see {Franx} 1993).

In star forming disks
aging of the stellar population is compensated by the formation of
new stars, and models predict only modest or even negative evolution in the
$M/L$ ratio with time ~(e.g., {Pozzetti}, {Bruzual A.}, \&  {Zamorani} 1996; {Ferreras} \& {Silk} 2001). The observed
evolution of the Tully-Fisher relation lends support to
these models, ruling out strong (1--2 mag) evolution over the redshift range
$0 < z < 1.3$, at least for the most massive galaxies
~({Vogt} {et~al.} 1996, 1997; {van Dokkum} \& {Stanford} 2001; {Ziegler} {et~al.} 2002).

By contrast, the $M/L$ ratios of
early-type galaxies (and the bulges of spiral galaxies)
are expected to
increase over time, as the demise
of massive stars causes their
stellar populations to fade ~({Tinsley} \& {Gunn} 1976).
The evolution of $M/L$ ratios of early-type galaxies has been measured
in rich clusters at $0.02\leq z \leq 0.83$
~(e.g., {van Dokkum} \& {Franx} 1996; {Kelson} {et~al.} 1997; {Bender} {et~al.} 1998; {van Dokkum} {et~al.} 1998),
and in the general field to $z\approx 0.7$
~({Treu} {et~al.} 1999; {van Dokkum} {et~al.} 2001a; {Treu} {et~al.} 2002),
using the Fundamental Plane relation.
The cluster data show a gradual increase of the mean $M/L_B$
ratio by a factor $\sim 3$ 
since $z=0.83$
(for $\Omega_m=0.3$, $\Omega_{\Lambda}=0.7$). The results for the general
field are still somewhat uncertain, but indicate that field early-type
galaxies have similar $M/L$ ratios as those in clusters, at least
out to $z\approx 0.6$.

In this paper, we present results on
the Fundamental Plane and $M/L$ ratios in the cluster \clusa\
at $z=1.27$ ~({Stanford} {et~al.} 1997), probably the highest redshift accessible 
for velocity dispersion measurements with
the current generation of telescopes. Although it is hazardous to extrapolate
the results presented here to the general population of distant red 
galaxies, studies of the Tully-Fisher relation and the FP at $z \gtrsim 1$
can be regarded as tentative first steps toward an empirical
determination of the mass function at high redshift.
We assume $\Omega_m=0.3$ and $\Omega_{\Lambda}=0.7$ throughout.
We used $H_0 = 50$\,\kms\,Mpc$^{-1}$ where needed, but note that our
results are not dependent on the value of the Hubble constant.

\section{Spectroscopy}

\subsection{Sample Selection and Observations}

The sample selection was based on an extensive
$BRIzJK_s$ survey of a 28 arcmin$^2$ field in Lynx
(Eisenhardt et al., in preperation). Keck spectroscopy and Chandra
imaging ~({Stanford} {et~al.} 2001) has revealed that the Lynx field
contains three distant X-ray clusters: \clusc\ at $z=0.57$
~({Holden} {et~al.} 2001), \clusb\ at $z=1.26$ ~({Rosati} {et~al.} 1999),
and \clusa\ at $z=1.27$ ~({Stanford} {et~al.} 1997).
Galaxies in the Lynx field
were selected on the basis of their $K_s$ magnitude.
Three multislit masks were designed. The primary sample
consists of seven $K_s<19$ galaxies with spectroscopic
redshifts $1.25 < z < 1.29$; these galaxies were repeated in each mask.
Remaining space
in the three masks was filled with $K_s<20$ galaxies without
spectroscopic redshift.

The Lynx field was observed on 2001 January 20--21 with the Low
Resolution Imaging Spectrograph ~(LRIS; {Oke} {et~al.} 1995) on the
Keck II Telescope. The D680 dichroic was used, in conjunction
with the 600\,lines\,mm$^{-1}$ grating blazed at
$1\mu$m in the red and the 300\,lines\,mm$^{-1}$ grism blazed at
5000\,\AA\ in the blue. The $1\farcs 2$ wide slits give a
resolution of 5.7\,\AA\ FWHM as measured from the width of atmospheric
emission lines, which corresponds to
$\sigma_{\rm instr} \approx 80$\,\kms\ at 9000\,\AA.
 Conditions were photometric, and the seeing was
$\approx 0 \farcs 9$. In order to facilitate sky subtraction the
galaxies were moved along the slit between successive exposures,
with offsets $-1\farcs 5$, $-4\farcs 5$, $+1\farcs 5$,
$+4\farcs 5$ with respect to the initial position. The integration
time for each exposure was 1800\,s. Each of the three masks
was observed for a total of 14.4\,ks, split in two sequences
of the four dither positions. The total integration time for
galaxies in the primary sample is 43.2\,ks, or twelve hours.

Initial results from our deep spectroscopy were presented in
~{van Dokkum} \& {Stanford} (2001). The present paper discusses the
three galaxies in the primary sample that fall within the area
of our deep {\em Hubble Space Telescope}
({\em HST}) WFPC2 and NICMOS imaging of \clusa\
at $z=1.27$. 
They are the brightest, second brightest,
and fourth brightest early-type galaxies in our $K$-selected
sample of cluster galaxies observed with {\em HST} ~({van Dokkum} {et~al.} 2001b).

\subsection{Reduction}

Each set of four dithered 1800\,s exposures was reduced separately.
Each slitlet was treated as a separate, long slit spectrum.
Bias was subtracted by fitting low order polynomials to the overscan
regions. The red CCD was read out using two amplifiers; the overscan region
for each amplifier was fitted separately. 
No systematic variations are present in the residuals
of bias frames
or the overscan regions of science images.

The most critical step
in the data reduction is the removal of the fringe pattern
in the red exposures.
Internal flatfields were taken every $\sim 90$
minutes, such that there was a flatfield within one hour
of each science exposure. For each slit, the flatfield was divided by
the average response in the wavelength direction. The response
was obtained by averaging the flatfield in the spatial direction
and fitting a fifth order polynomial in the wavelength direction.
Flatfielding reduced the peak-to-peak variation in the fringe
pattern from $\approx 7$\,\% to
$\approx 2$\,\%.
A map of the residual fringe pattern was created from
the four dithered exposures in the following way.
First, sky lines were fitted by a
third order polynomial, masking the galaxy spectrum (and
any other objects) in the fit. Next, the median of the four
sky subtracted spectra was computed, disregarding pixels at
the positions of object spectra.   Finally,
this median residual map
was fitted by a low order polynomial in the wavelength direction
to remove faint unmasked objects and any residuals from galaxy spectra.
The fringe map was subtracted from each of
the four exposures. Residual peak-to-peak
variations are $< 0.5$\,\%, and the
noise in the final combined spectra is dominated by
Poisson fluctuations.

Cosmic ray removal is not straightforward, despite the large number of
independent exposures. As a result of our dithering procedure
and flexure in the spectrograph
the positions of galaxy spectra as well as those of sky lines
change from one exposure to the next.
Therefore, the sky lines and object spectra were modeled and subtracted
before identifying cosmic rays in the residual images. Sky lines
were modeled by fitting low order polynomials. The object spectrum was
modeled by taking the median of the four object
spectra extracted from the individual dithered
exposures, and smoothing with a $3 \times 3$
boxcar filter. The model sky and the model galaxy (shifted appropriately
for each dithered exposure) were summed to create 2D model spectra,
which were subtracted from each exposure.
Cosmic rays were identified by comparing the flux in the residual
images to the expected noise calculated from the 2D model.
Pixels affected by cosmic rays were replaced by the
values of corresponding pixels in the model.

Sky lines were subtracted from the cosmic ray cleaned
exposures by fitting a third order polynomial
in the spatial direction, masking the galaxy and any other
objects in the slit. Subtracting the sky lines before wavelength
calibration and rectification has the disadvantage that the
lines are still slightly tilted. However, the advantage of this procedure
is that aliasing due to rectification of sharp, bright
lines is avoided; tests showed that such aliasing effects
can be avoided entirely
by reversing the usual reduction procedure.

A separate wavelength solution was obtained for each of the four
galaxy spectra.  Arc lamp exposures were taken at regular intervals
during the night.  A fit to the positions of bright lines in the
nearest arc exposure provided the initial solution. The observed
locations of the [O\,{\sc i}]~5577.34\,\AA\ sky line in the blue and
the OH~7--3~P1~8885.85\,\AA\ sky line in the red were used to correct
the zeropoint for each spectrum; these corrections were typically
$\sim 3$\,\AA. The sky subtracted
2D spectra were transformed to a common log\,$\lambda$ scale.

The galaxy spectra are too faint in individual 1800\,s exposures
for a reliable S-distortion correction. Therefore, the four spectra
were first combined into a single 2D spectrum, using linear weighting by
the signal-to-noise (S/N)
ratio. The S-distortion was removed by binning the
spectrum in 100\,\AA\ wide bins and fitting a low order polynomial.
Atmospheric absorption features were removed using the spectrum of
a bright blue star that was included in each of the masks.
A long slit spectrum of the star G191B2B ~({Massey} \& {Gronwall} 1990)
taken with the same
instrumental setup provided an approximate
flux calibration. Finally, the six 7200\,s spectra
were combined into a single 43.2\,ks spectrum, using optimal weighting.

\subsection{Spectral Properties}
\label{spec.sec}

\begin{figure*}[t]
\begin{center}
\leavevmode
\hbox{%
\epsfxsize=16cm
\epsffile{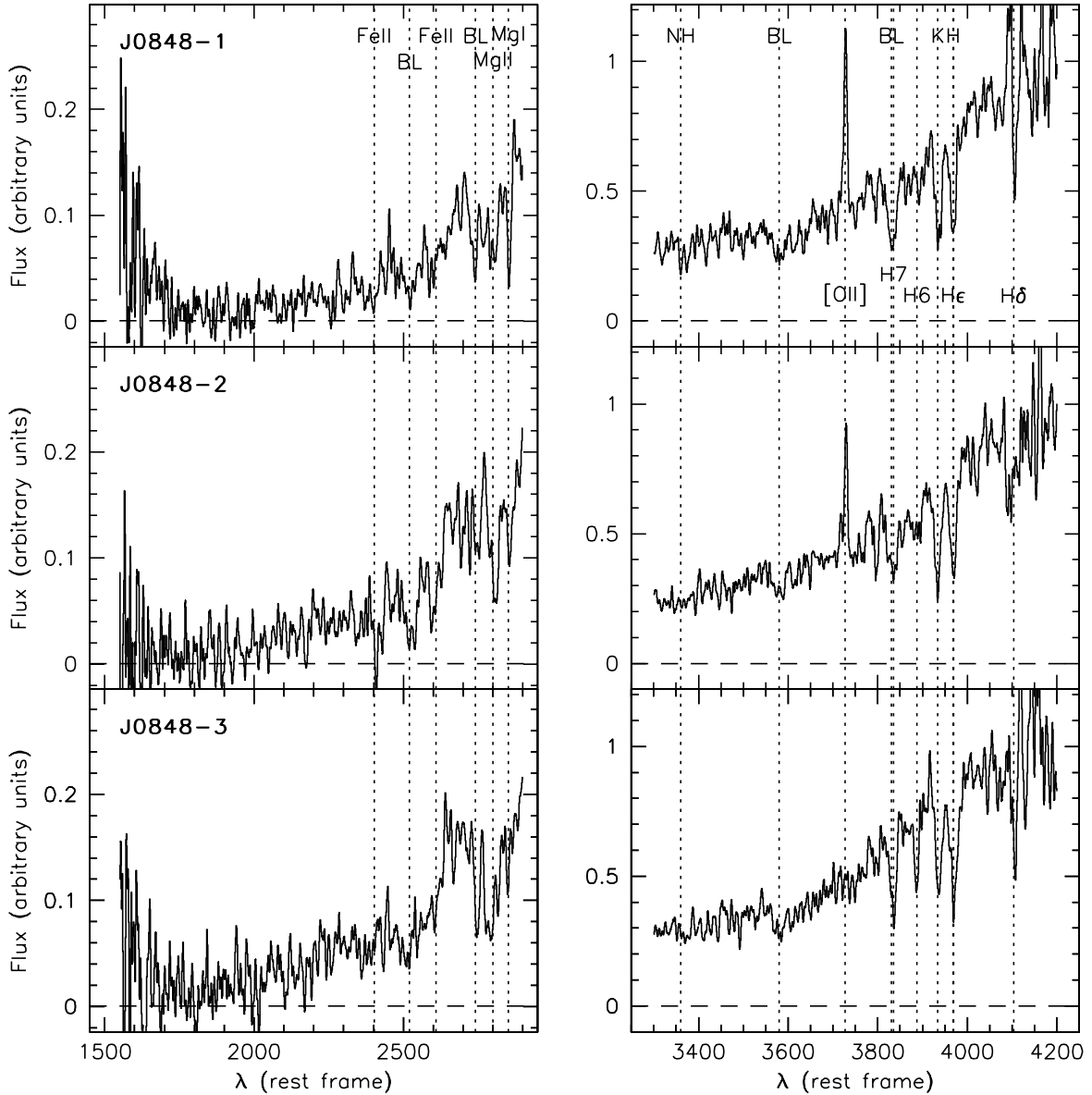}}
\figcaption{
\small
Keck spectra of three early-type galaxies in the $z=1.27$ cluster
\clusa, smoothed by a boxcar filter of width
10\,\AA\ in the observed frame.
Spectra from the blue arm and
the red arm of LRIS are shown separately. The exposure time is 43.2\,ks.
Galaxies \gala\ and \galb\ show \oii\ emission; galaxy
0848--3 has strong Balmer absorption lines.
\label{full.plot}}
\end{center}
\end{figure*}

Spectra of the three galaxies that are covered by
our {\em HST} imaging are shown in Fig.\ \ref{full.plot},
smoothed with a boxcar filter of width 10\,\AA\
in the observed frame to enhance broad absorption features.
The 1D spectra were extracted by summing the five central
rows ($1\farcs 1$) in the 2D spectra. No weighting was applied, to
facilitate aperture corrections (see \S~\ref{sigma.sec}).
For each galaxy the S/N per \AA\ at $\lambda_{\rm rest}=3700$\,\AA\
(calculated from the unsmoothed data) is listed in Table 1.

The most prominent absorption features are \hk,
Mg\,{\sc i} and Mg\,{\sc ii},
and the $\sim 2600$\,\AA\ break.
Although evolved stellar populations
have many absorption features
between $\sim\,3400$\,\AA\ and $\sim 3900$\,\AA,
line indices have been
defined for only a handful
~(see {Ponder} {et~al.} 1998). The strongest are two intrinsically
broad lines: a blend at 3580\,\AA\ ~({Davidge} \& {Clark} 1994)
and a blend of Fe\,I and Mg\,I at $3840$\,\AA\ 
~({Pickles} 1985).

Interestingly, galaxies \gala\ and \galb\ show \3727\ emission,
with rest-frame equivalent width $-17 \pm 2$\,\AA\ and $-13 \pm 2$\,\AA\
respectively.
Furthermore,
all three galaxies show some evidence for enhanced H$\delta$ absorption,
although measurements in this spectral region are severely affected
by sky line residuals. In galaxy \galc\ the higher order
Balmer lines H$\epsilon$, H6, and H7
are clearly enhanced as well. H$\epsilon$ and H7 are blended with
Ca\,H and Fe\,I + Mg\,I
at $3840$\,\AA\ respectively, but H6 is not contaminated by
any strong metal lines. The rest-frame equivalent width of H6
is $4.5 \pm 0.8$\,\AA, making \galc\ the highest redshift
``E+A'' galaxy ~({Dressler} \& {Gunn} 1983) currently known.

No UV upturn is evident at $\lambda \lesssim 2500$\,\AA,
consistent with the idea that hot horizontal branch stars
do not contribute significantly to the mid-UV flux
in stellar population with ages less than
$\sim 5$ Gyr ~(e.g., {Dorman}, {O'Connell}, \& {Rood} 1995; {Brown} {et~al.} 2000).
A full analysis of the mid-UV spectral features is beyond the scope of
the present paper; we note here that
the mid-UV spectra are very similar to
those of nearby galaxies having
intermediate age populations, such as NGC\,3610 or M32
~(see {Lotz}, {Ferguson}, \& {Bohlin} 2000).

\subsection{Velocity Dispersions}
\label{sigma.sec}

Kinematics of nearby
early-type galaxies are usually determined from the
spectral region containing the strong \mgb\ line at 5172\,\AA\
~(e.g., {Davies} {et~al.} 1987; {Lucey} {et~al.} 1991; {J\o{}rgensen}, {Franx},  \& {Kj\ae{}rgaard} 1995b; {Mehlert} {et~al.} 2000).
Studies of early-type
galaxies at $z\gtrsim 0.5$ have used bluer spectral regions
near the G-band at 4300\,\AA\
~(e.g., {van Dokkum} \& {Franx} 1996; {Kelson} {et~al.} 2000a; {Koopmans} \& {Treu} 2002).
Beyond $z \approx 1.1$ it
seems attractive to use the strong
\hk\ lines for measuring velocity dispersions
~(see, e.g., {Kobulnicky} \& {Gebhardt} 2000). However, because
these lines are intrinsically broad
($\sigma \approx 5$\,\AA) they are not very sensitive indicators
of velocity dispersion. Additional problems are template
mismatch caused by errors in the continuum fitting
across the 4000\,\AA\ break and the fact that the Ca\,{\sc ii} H
line at 3968\,\AA\ is blended with H$\epsilon$ at 3969\,\AA.
Empirically,
the \hk\ lines appear to overestimate the velocity dispersion
~({Kormendy} \& {Illingworth} 1982).

In the present study we explore the use of near-UV lines blueward of
the \hk\ lines for measuring velocity dispersions of high redshift
galaxies. Central velocity dispersions were determined from a fit to a
convolved template star spectrum in real space, following the
procedures outlined in ~{van Dokkum} \& {Franx} (1996) and ~{Kelson}
{et~al.} (2000a). The wavelength region is 3400 -- 4000\,\AA\ in the
rest-frame, with regions around \3727\ and the Balmer lines
masked. The spectra were weighted by the S/N ratio, as determined from
sky spectra. 

The solar spectrum\footnote{Obtained from the BAse Solaire Sol 2000
(http://mesola.obspm.fr/form\_spectre.html).} was used in the fits,
because it spans the entire rest-frame wavelength range of our
LRIS-R observations at high S/N ratio, and its spectral type
is expected to be appropriate for the integrated light of intermediate
age stellar populations. The validity of this approach is tested
in Sect.\ \ref{nearby.sec} on two nearby galaxies.
For each galaxy the wavelength-dependent
instrumental resolution was determined from sky lines, and the solar
spectrum was smoothed
and rebinned to match the resolution of the galaxy spectra.

Fits and their residuals are shown in Fig.\ \ref{red.plot},
and the resulting velocity dispersions are listed in Table 1.
For consistency with earlier work
the dispersions in our
$0\farcs 9 \times 1\farcs 1$ rectangular apertures
were corrected to a circular aperture of $3\farcs 4$
diameter at the distance of Coma, following the procedure of
~{J\o{}rgensen} {et~al.} (1995b).
The errors in Table 1 do not include a systematic uncertainty
of $\sim 5$\,\%, determined from varying
the continuum filtering, the wavelength range,
and the weighting scheme. Limiting the fitting range
to $3400 - 3900$\,\AA\ had a negligible effect on the dispersions
and their uncertainties, and no stable solutions were found when
only the region $3800 - 4000$\,\AA\ was used in the fit.
These results suggest that the \hk\ lines have little weight in
the fit, as expected from their large intrinsic width.

We were unable to measure stable dispersions when we
limited the wavelength range to $3400 - 3700$\,\AA, or split
the data in two independent datasets with effective
exposure time 21.6\,ks. These tests indicate that our observations
just reach the required S/N for velocity dispersion measurements at
this redshift.\vspace{-1cm}\\

\begin{small}
\begin{center}
{ {\sc TABLE 1} \\
\sc Velocity Dispersions} \\
\vspace{0.1cm}
\begin{tabular}{lccc}
\hline
\hline
Galaxy & S/N & $z$ &  $\sigma$\,(\kms) \\
\hline
\gala\ & 14 & $1.2751$ & $237 \pm 34$ \\
\galb\ & 15 & $1.2770$ & $308 \pm 41$ \\
\galc\ & 17 & $1.2774$ & $174 \pm 29$ \\
\hline
\end{tabular}
\end{center}
\end{small}

\null
\vbox{
\begin{center}
\leavevmode
\hbox{%
\epsfxsize=8.2cm
\epsffile{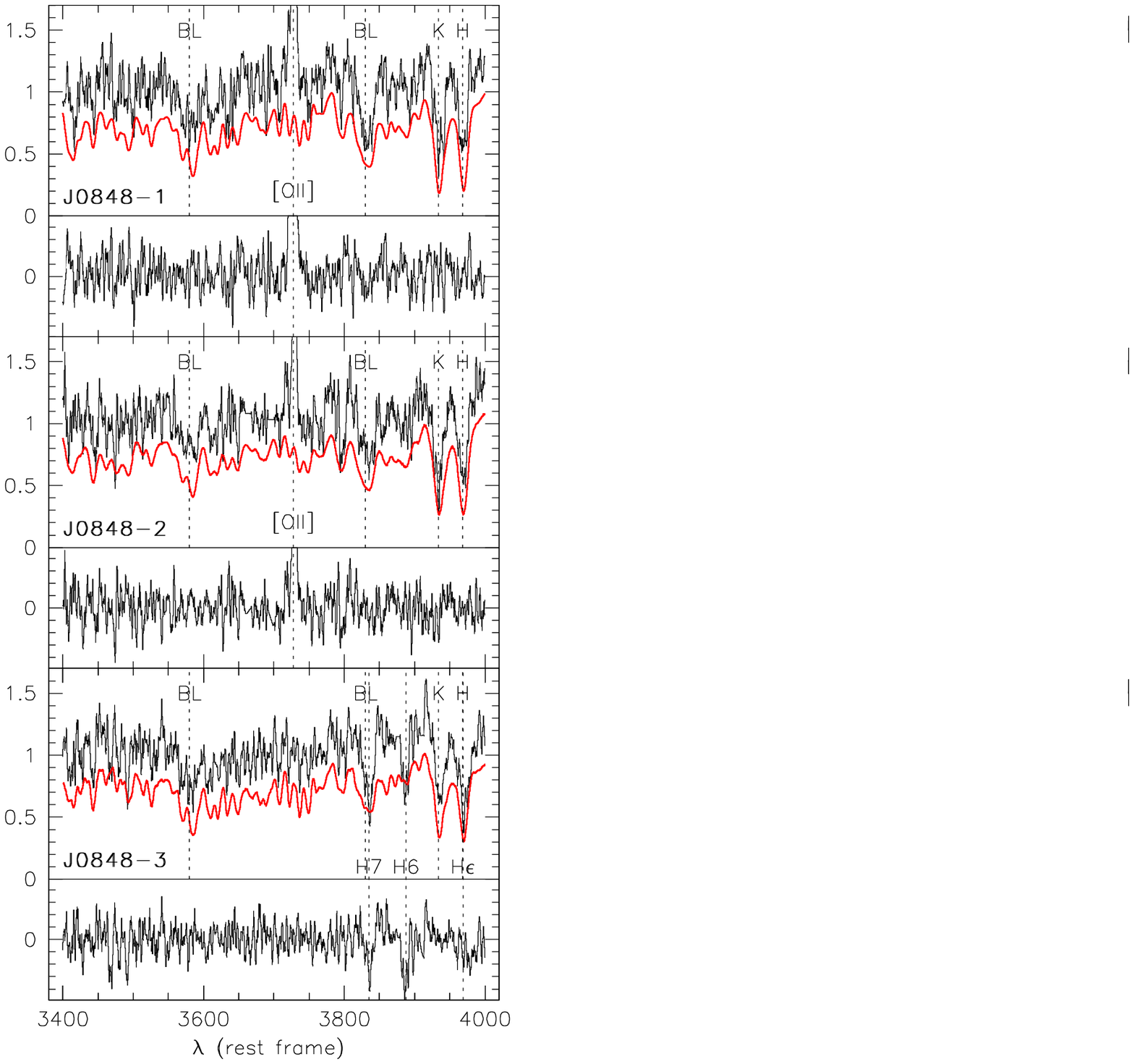}}
\figcaption{\small
The spectral region 3400 -- 4000\,\AA\ at full resolution, i.e.,
without any smoothing. The solar
spectrum is shown in red, smoothed to the velocity dispersion of the
galaxies, and offset for clarity. Below each spectrum the residuals
from the fit are shown. Note the higher order Balmer lines in the
residuals of \galc.
\label{red.plot}}
\end{center}}

\subsection{Tests on Two Nearby Galaxies}
\label{nearby.sec}

Moderate resolution near-UV spectra were obtained for two nearby
galaxies, NGC~3610 and NGC~5018, in order to test whether velocity
dispersions obtained from near-UV lines are unbiased with respect to
determinations from redder spectral regions.  The galaxies were
selected because they are believed to have comparitively young
luminosity weighted ages; galaxies at $z=1.27$ are at most $\sim
5$ Gyr old, and their stellar populations may resemble those of young
early-type galaxies in the local Universe (see Sect.\ \ref{spec.sec}).

NGC~3610 is classified
as E5, has significant morphological fine structure indicating
a recent merger, and is thought to have
a centrally concentrated intermediate-age stellar population
~({Silva} \& {Bothun} 1998, and references therein). The galaxy also has
a dynamically cold disk ~({Rix} \& {White} 1992).
NGC~5018 is an elliptical galaxy with shells,
and is also considered to be a recent merger remnant
~(e.g., {Fort} {et~al.} 1986). Modelling of line indices
indicates the presence of a $\sim 3$\,Gyr old stellar population
dominating the light at $\sim 4000$\,\AA\ ~({Leonardi} \& {Worthey} 2000).

The galaxies were observed 2001 June 19, with the LRIS Double
Spectrograph on the Keck I Telescope, using a $1\farcs 0$ slit
and the D460 dichroic. The two arms of LRIS allow simultaneous
measurement of velocity dispersions in the red and the near-UV,
sampling approximately the same spatial region of the galaxies.
The blue arm used the
1200\,lines\,mm$^{-1}$ grism blazed at 3400\,\AA,
giving a wavelength range of $3000$\,\AA\
-- $3800$\,\AA. The red arm used the 900\,lines\,mm$^{-1}$
grating blazed at 5500\,\AA; the grating angle that
was used gives a wavelength range of $4700$
-- $5950$\,\AA.
The instrumental resolution as measured
from night sky lines $\sigma_{\rm instr} \approx 65$\,\kms\
at 3500\,\AA\ and $\approx 70$\,\kms\ at 5200\,\AA.
Exposure times were 1800\,s for NGC~3610 and
1800\,s for NGC~5018. The observations were done through cirrus
and occasional clouds. Spectra of template stars HD\,102494
(G9IV),
HD\,132737 (K0III), and HD\,210220 (G6III) were obtained in twilight, using
the same instrumental setup.

The reduction followed standard procedures for long slit spectroscopic
data. Bias levels were determined from the overscan regions on the
CCD. Internal flat fields were used to correct for the pixel-to-pixel
variation of the CCD response.  Cosmic rays were removed using the
{\sc L.A.Cosmic} task ~({van Dokkum} 2001).  Arc lamp lines were used
for wavelength calibration, with zeropoint offsets determined from
sky lines in the object spectra.  Sky spectra were determined
from the edges of the slit, and subtracted.
Normalized near-UV spectra of
NGC~3610 and NGC~5018 are shown in Fig.\ \ref{lowz.plot}. 
The 1D spectra were
extracted by summing ten rows, corresponding to $2\farcs 2$.
The S/N ratio per \AA\ is 55 for
NGC~3610 and 23 for NGC~5018. The region
3465\,\AA\ -- 3495\,\AA\ is not shown because the galaxy spectra are
severely affected by the presence of an undispersed slit image in that
region, caused by insufficient baffling of the blue grism.

Velocity dispersions were determined from a direct fit of the template
stars to the galaxy spectra, as discussed in Sect.\ \ref{sigma.sec}.
In addition to the
three template stars observed with LRIS-B the
solar spectrum was used in the fits.
The solar spectrum was smoothed and rebinned to match the resolution
of the galaxy spectra. The results are listed in Table 2.

There is evidence that the red spectral region gives slightly
higher velocity dispersions (by $\sim 5$\,\%) than the near-UV
region. If the effect is real, it
may indicate a bias in the measurement technique, or reflect
the presence of a dynamically cold component
of young stars in these two galaxies.
The small difference may also be caused
by differential atmospheric refraction (causing misalignment of the slit
and the galaxy center in the near-UV), since both galaxies
were observed at high airmass.
The sign of the difference is
consistent with this explanation, but the effect
is difficult to quantify
because early-type galaxies often show complex structure
in their central velocity dispersion profiles (e.g., van der Marel
et al.\ 1994).

\vbox{
\begin{center}
\leavevmode
\hbox{%
\epsfxsize=8.2cm
\epsffile{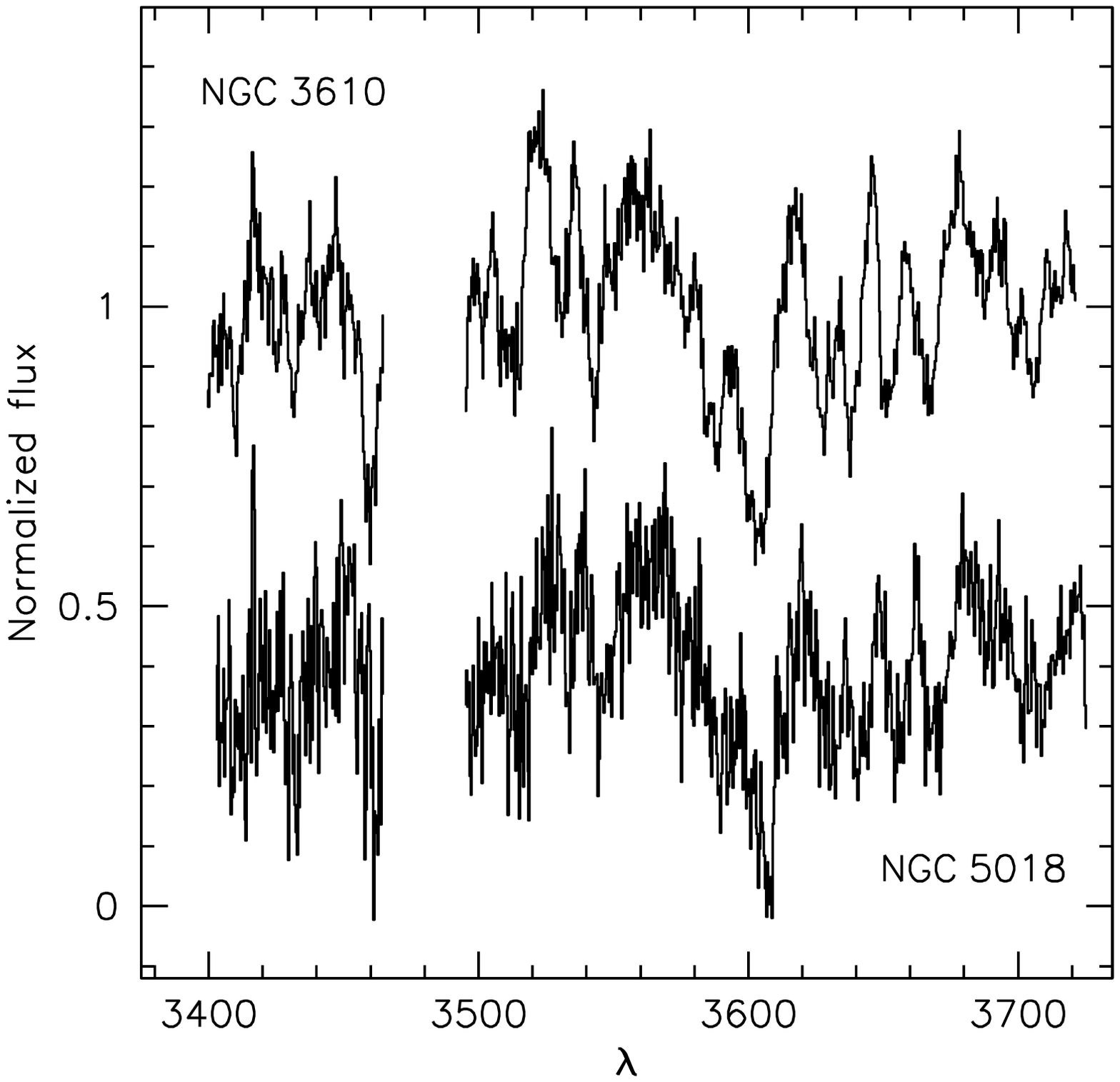}}
\figcaption{\small
Near-UV spectra of NGC~3610 (top) and NGC~5018 (bottom),
at a resolution
$\sigma_{\rm instr} \approx 65$\,\kms. Both spectra were normalized
by fitting a low order polynomial; for clarity the spectrum of NGC~5018
was offset by $-0.6$. Many moderately strong absorption lines
are present in this spectral region.
The spectra are significantly affected by stray light in the
region 3465\,\AA\ -- 3495\,\AA, which is excluded in the velocity
dispersion fitting.
\label{lowz.plot}}
\end{center}}

The Sun gives very similar results as the other template
stars. We conclude that determining velocity dispersions from
the near-UV lines using the solar spectrum as a template may introduce
systematic errors of $\sim 5$\,\%. Assuming the systematic
errors can be added quadratically, we find a total
systematic uncertainty of $\sim 7$\,\%.

\begin{small}
\begin{center}
{ {\sc TABLE 2} \\
\sc Velocity Dispersions of Nearby Galaxies} \\
\vspace{0.1cm}
\begin{tabular}{llcc}
\hline
\hline
Galaxy & Template &  $\sigma$ (Mg$_b$)& $\sigma$ (3600) \\
\hline
NGC~3610 & HD\,102494 & $143\pm 2$  & $134\pm 6$ \\
NGC~3610 & HD\,132737 & $143\pm 2$  & $134\pm 5$ \\
NGC~3610 & HD\,210220 & $143\pm 2$  & $134\pm 5$ \\
NGC~3610 & Sun & $144\pm 2$ & $135\pm 6$ \\
NGC~5018 & HD\,102494 & $206\pm 3$& $199\pm 15$  \\
NGC~5018 & HD\,132747 & $207\pm 3$ & $197\pm 15$  \\
NGC~5018 & HD\,210220 & $206\pm 3$ & $199\pm 15$  \\
NGC~5018 & Sun & $206\pm  3$ & $194\pm 15$ \\
\hline
\end{tabular}
\end{center}
\end{small}

\section{Photometry}

\begin{figure*}[p]
\begin{center}
\leavevmode
\hbox{%
\epsfxsize=17cm
\epsffile{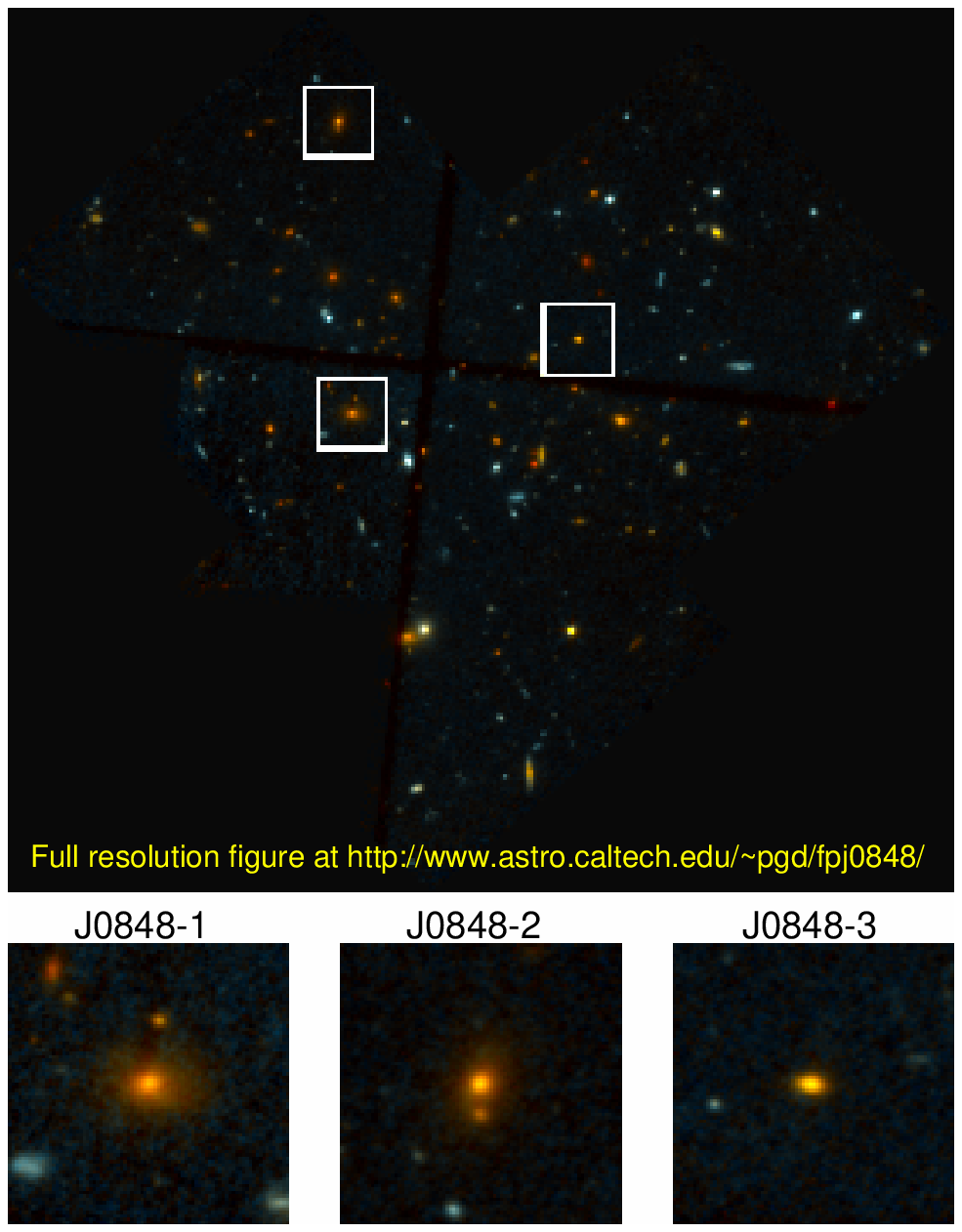}}
\figcaption{
\small
Color rendition of the core of \clusa\ at $z=1.27$, created from
a WFPC2 $I_{F814W}$ image and a mosaic of three {\em HST} NICMOS $H_{F160W}$
images. The WFPC2 data were smoothed and resampled to match the
resolution of the NICMOS data. Most of the objects appearing orange
are cluster members. The insets show the three early-type galaxies
with measured velocity dispersions. The scale of the large image is
$2\farcm 23 \times 2 \farcm 23$, and of the insets $10\arcsec
\times 10\arcsec$.
\label{image.plot}}
\end{center}
\end{figure*}

Colors and structural parameters were determined from
{\em HST} imaging previously described in ~{van Dokkum} {et~al.} (2001b).
The cluster was observed in the $I_{F814W}$ band with the Wide
Field and Planetary Camera 2 (WFPC2),
and in the $H_{F160W}$ band with the Near Infra-Red Camera and
Multi-Object Spectrometer (NICMOS) NIC3 camera.
We refer to ~{van Dokkum} {et~al.} (2001b) for details of the observations
and reduction. Figure \ref{image.plot} shows a
color image of \clusa\ created from our {\em HST} imaging. The pixel
scale of our final drizzled images is $0\farcs 04$; the
total area covered by both WFPC2 and NICMOS is 2.3 arcmin$^2$.

\begin{figure*}[t]
\begin{center}
\leavevmode
\hbox{%
\epsfxsize=18cm
\epsffile{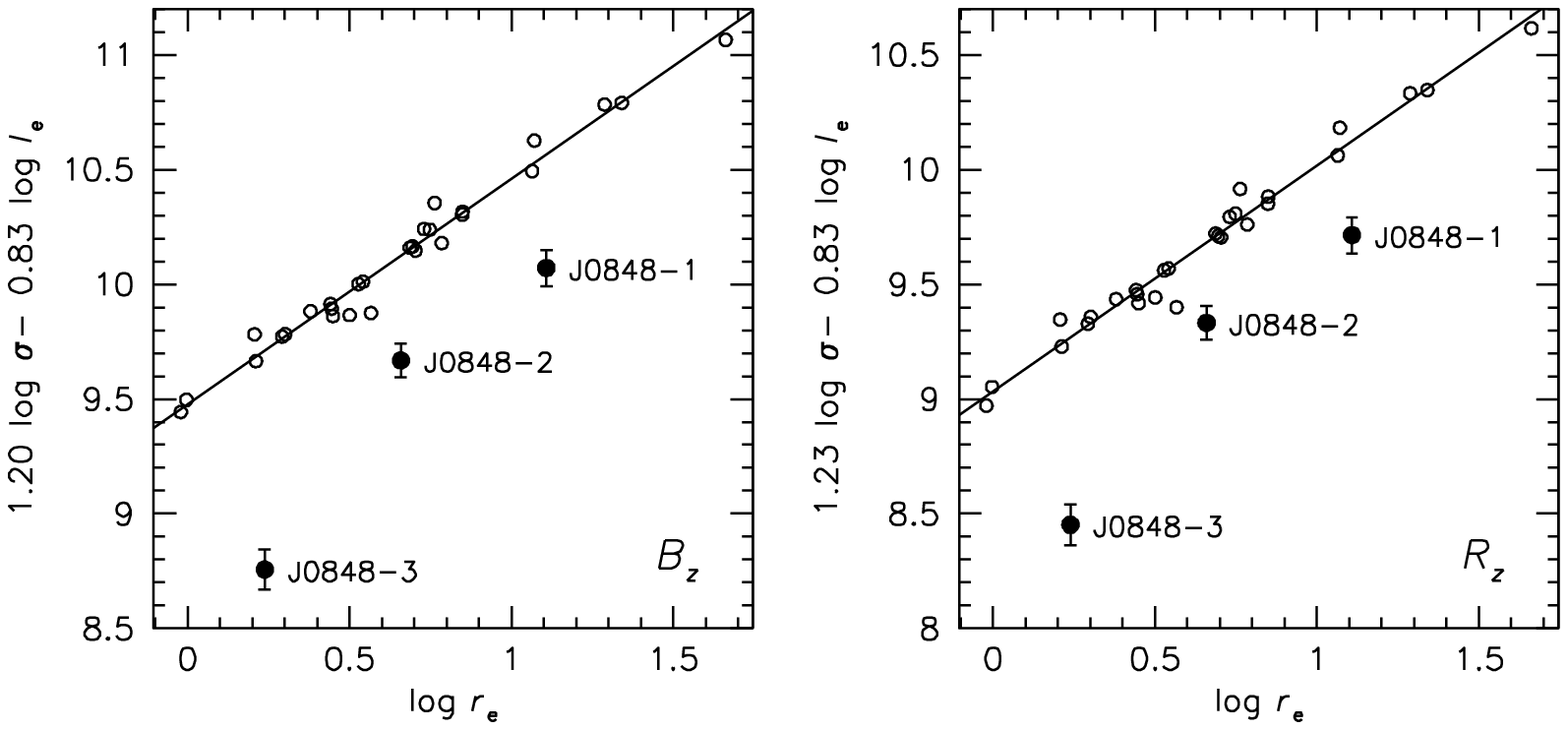}}
\figcaption{
\small
Edge-on projection of the Fundamental Plane of the nearby Coma
cluster (J\o{}rgensen et al.\ 1996)
in the $B$ and $R$ bands (open symbols).
Solid symbols show the three early-type
galaxies at $z=1.27$.
Surface brightnesses were corrected for
$(1+z)^4$ cosmological dimming, and transformed to the rest-frame
$B$ and $R$ bands; $\log I_e \equiv -0.4 \mu_e$.
The $z=1.27$ galaxies are offset with respect
to the FP of Coma due to passive evolution of the stellar populations.
The offsets are larger in the
rest-frame $B$ band than in the rest-frame $R$ band.
In both bands the offset is greatest for the ``E+A'' galaxy
\galc. More data are needed to determine the form of the FP at
$z=1.27$.
\label{fp.plot}}
\end{center}
\end{figure*}

\subsection{Colors}

Colors were measured from the $H_{F160W}$ and $I_{F814W}$
images in $1\farcs 5$ diameter apertures, after smoothing
the $I_{F814W}$ images to the NICMOS resolution. Zeropoints
on the Vega system
were obtained from the Data Handbooks (STScI, Baltimore) of
NICMOS and WFPC2 respectively:
\begin{eqnarray}
H_{F160W} & = & -2.5 \log ({\rm ADU}/s) + 21.496\\
I_{F814W} & = &  -2.5 \log ({\rm ADU}/s) + 21.688.
\end{eqnarray}
A correction of 0.05 magnitudes was applied to the $I_{F814W}$
zeropoint to account for the long/short anomaly
(see, e.g., Wiggs et al.\ 1998). Galactic extinction
is very small in this field
~($E(B-V) \approx 0.02$; {Schlegel}, {Finkbeiner}, \&  {Davis} 1998) and
was ignored. Measured colors are listed in Table 3. Errors are
estimated at $0.07$, dominated by systematic uncertainties
in the zeropoints.

\begin{small}
\begin{center}
{ {\sc TABLE 3} \\
\sc Photometry} \\
\vspace{0.1cm}
\begin{tabular}{lccccc}
\hline
\hline
Galaxy & $K_s$ & $I_{814}$ & $I_{814}-H_{160}$ & $\log r_e$ & $\mu_{e,160}$ \\
       & Total & $r<0\farcs 75$  & $r<0\farcs 75$& $\arcsec$ &  \\
\hline
\gala\ & 17.36 & 22.90 & 3.48 & $-0.010$ & $21.75$ \\
\galb\ & 17.97 & 22.94 & 3.35 & $-0.460$ & $20.19$ \\
\galc\ & 18.71 & 22.85 & 3.10 & $-0.880$ & $18.49$ \\
\hline
\end{tabular}
\end{center}
\end{small}

\subsection{Structural Parameters}

Following the procedures outlined in ~{van Dokkum} \& {Franx} (1996) effective radii
and effective surface brightnesses were determined for the three galaxies
with measured velocity dispersions.  For each galaxy, models
consisting of an $r^{1/4}$ law convolved with a Point Spread Function
(PSF) were fitted directly to the two-dimensional images. The NICMOS
$H_{F160W}$ images were used in the fits, because the galaxies have
much higher S/N than in the WFPC2 $I_{F814W}$ images. Furthermore
effects of color gradients are minimized by fitting in the rest-frame
$R$-band rather than the rest-frame $U$-band.

A separate PSF was used for each galaxy. The PSFs were created by
applying the same dither pattern as used in the observations
to model PSFs generated by Tiny Tim ~({Krist} 1995).
The PSFs were then combined and rebinned using ``drizzle''
~({Fruchter} \& {Hook} 1997), in the same way as the
observations. We note that
our results are not very sensitive to errors in the PSFs
or color gradients, because
the error in the product $r_e I_e^{0.83}$ is
almost parallel to the FP ~(see, e.g., {J\o{}rgensen}, {Franx},  \& {Kj\ae{}rgaard} 1995a).
Results from the fits are listed in Table 3.
Effective radii are in arcseconds;
surface brightnesses are in $H_{F160W}$ mag/arcsec$^2$ and
not corrected for $(1+z)^4$ cosmological dimming.

For a meaningful comparison between galaxies at different
redshifts,  measured surface brightnesses need to be converted to a
common rest-frame band. We assume that the flux density in the
rest-frame $B$ band can be related to the flux density
in the observed $I_{F814W}$ and $H_{F160W}$ bands by
$F_{B(z)} = F_I^{\alpha} F_H^{1-\alpha}$
~(see {van Dokkum} \& {Franx} 1996). Using
spectral energy distributions from ~{Coleman}, {Wu}, \& {Weedman} (1980) we find
for $z=1.27$
\begin{equation}
\label{bz.eq}
B_z = H_{F160W} + 0.50 (I_{F814W} - H_{F160W}) + 1.81
\end{equation}
and, similarly,
\begin{equation}
\label{rz.eq}
R_z = H_{F160W} + 0.02 (I_{F814W} - H_{F160W}) + 1.95,
\end{equation}
with the subscript $z$ denoting rest-frame (i.e., redshifted) band.
These transformations are very different from a traditional
$K$-correction, as they use the observed colors of the galaxies to
interpolate between passbands.  As a result, they are independent of
spectral type to $\sim 0.03$ mag.
The total systematic error is estimated at $\sim 0.05$, which
includes uncertainties in absolute spectrophotometry
(see, e.g., van Dokkum \& Franx 1996).
Note that the redshifted Cousins $R$ filter is a close match to the observed
$H_{F160W}$ filter.

As a test on our entire procedure we also determined structural parameters
from the WFPC2 $I_{F814W}$ images, using Tiny Tim PSFs and Eq.\ \ref{bz.eq}
to convert surface brightnesses to the $B_z$ band. 
We find that the difference in the
parameter $r_e I_e^{0.83}$ is $5$\,\% for \gala,
$8$\,\% for \galb, and $16$\,\% for \galc. Given the much higher S/N
in the NICMOS images these results can be viewed as
upper limits to the true uncertainties. We conclude that the
uncertainties in the $M/L$ ratios of individual galaxies are
dominated by random errors in the velocity dispersions.

\section{Fundamental Plane and Mass-to-Light Ratios}

The velocity dispersions and structural parameters allow us to
study the Fundamental Plane of early-type galaxies at $z=1.27$.
The Fundamental Plane of nearby clusters has the form
\begin{equation}
r_e \propto \sigma^{\alpha} I_e^{\beta}
\end{equation}
with $I_e$ surface brightness at the effective radius in linear units.
~{J\o{}rgensen}, {Franx}, \&  {Kj\ae{}rgaard} (1996) find $\alpha=1.20$, $\beta=-0.83$ in the $B$-band
and $\alpha = 1.23$, $\beta=-0.83$ in Gunn $r$ from an analysis
of 225 early-type galaxies in ten nearby clusters.
Assuming that early-type galaxies form a homologous family, the observed
tilt of the FP implies that the $M/L$ ratios of galaxies scale with
their mass as
\begin{equation}
\label{ml.eq}
M/L \propto M^{0.28}r_e^{-0.07}
\end{equation}
in the $B$ band. Therefore, the evolution of the
Fundamental Plane tracks the evolution of the $M/L$ ratios of galaxies
~({Franx} 1995).

\subsection{Edge-on Projection of the Fundamental Plane}

In Fig.\ \ref{fp.plot} open symbols
show the edge-on projection of the FP of the nearby
Coma cluster (J\o{}rgensen et al.\ 1996), and solid symbols show
the three galaxies in \clusa\ at $z=1.27$.
The J\o{}rgensen et al.\ Gunn $r$ data were transformed to
Cousins $R$ using $R-r=0.35$ (Fukugita, Shimasaku, \& Ichikawa 1995).
Surface brightnesses were corrected for $(1+z)^4$ cosmological dimming.
All three galaxies show a large offset from the FP of Coma,
and we conclude that the galaxies at $z=1.27$ do not occupy the
same Fundamental Plane as Coma galaxies.
The offsets are smaller in the $R_z$ band than in
the $B_z$ band, consistent with the expectation that
early-type galaxies at $z=1.27$ are bluer than those in Coma.
Galaxies \gala\ and \galb\
show a much smaller offset than the E+A galaxy \galc.

Our sample of three galaxies is obviously too small to 
determine the coefficients $\alpha$ and $\beta$.
The tilt of the FP may evolve with time
due to various processes, such as  systematic
age differences between low mass galaxies and high mass galaxies.
A study of 30 early-type galaxies in the cluster CL\,1358+62 at $z=0.33$
has shown that the form of the FP has not changed
significantly over the past $\sim 4$\,Gyr
~({Kelson} {et~al.} 2000b). At higher redshift the samples are still
too small to place strong constraints on the tilt.
In the following, we do not strictly assume that
the form of the FP at $z=1.27$ is the same as in nearby clusters, 
but we do assume that offsets from
the Coma FP are due to changes in $M/L$ ratio resulting from
stellar evolution. This interpretation is supported by the fact
that the E+A galaxy shows the largest deviation, and
it is consistent with
the reduced offsets in $R_z$ compared to $B_z$.

\begin{figure*}
\begin{center}
\leavevmode
\hbox{%
\epsfxsize=17cm
\epsffile{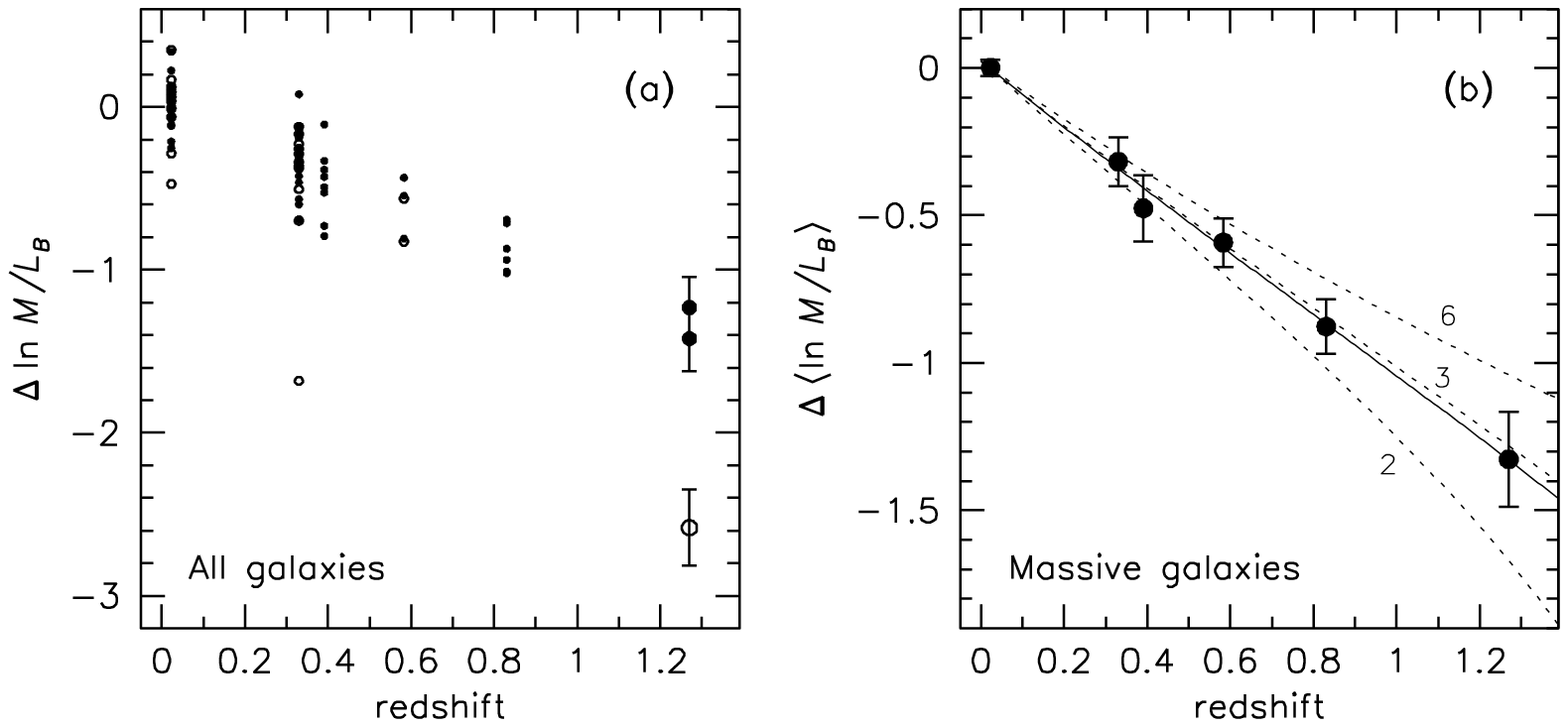}}
\figcaption{
\small
Evolution of the $M/L$ ratio of early-type cluster galaxies, in the
rest-frame $B$ band. In (a) individual galaxies in clusters
at $0.02\leq z \leq 1.27$ are shown. Data at $z<1$ are taken from 
J\o{}rgensen et al.\ 1996,
van Dokkum et al.\ (1998) and Kelson et al.\ (2000).
Open symbols are galaxies with masses $M
<10^{11}\,M_{\odot}$. In (b) the data for each cluster are averaged,
excluding galaxies with masses $M < 10^{11}\,M_{\odot}$.
Broken lines are
predictions from simple single burst stellar population synthesis models,
with $z_{\rm form} = 6$, $3$, and $2$. The solid line
is a prediction from models incorporating morphological evolution,
taken from van Dokkum \& Franx (2001).
The $M/L$ ratios of the two most
massive galaxies in \clusa\ are consistent with an extrapolation
of results obtained at $z<1$.
\label{res.plot}}
\end{center}
\end{figure*}

\vbox{
\begin{center}
\leavevmode
\hbox{%
\epsfxsize=8.4cm
\epsffile{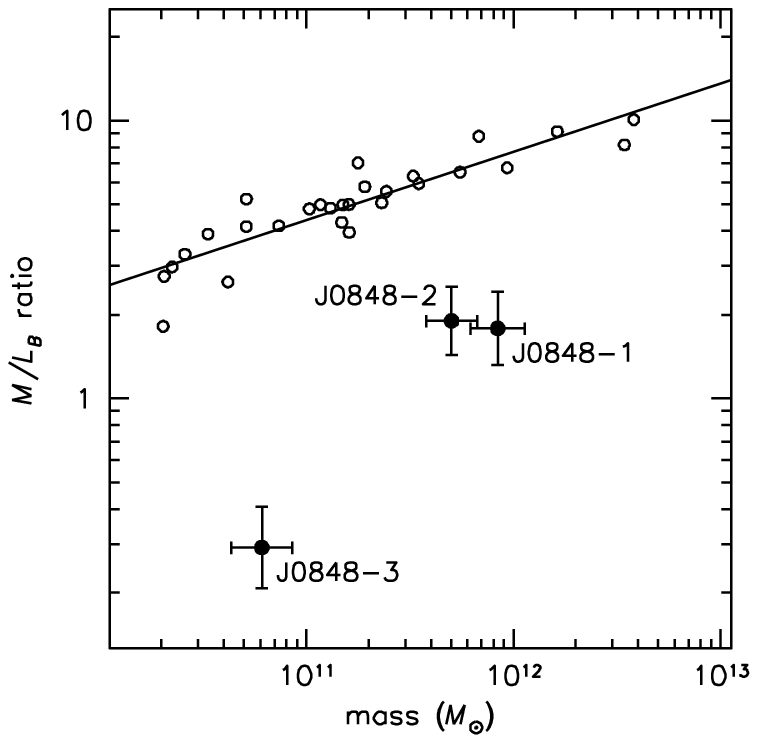}}
\figcaption{\small
Rest-frame $B$ band mass-to-light ratio versus mass, for galaxies in Coma
(open circles) and \clusa\ (solid circles). Galaxy \galc, the galaxy with
the largest offset from Coma (i.e., the youngest stellar population),
is a factor $\sim 10$ less massive than \gala\ and \galb.
\label{ml.plot}}
\end{center}}

\subsection{Evolution of Mass-to-Light Ratios}

The evolution of the mean $M/L$ ratio can be determined directly from
the evolution of the zeropoint of the FP (see Eq.\ \ref{ml.eq} and
van Dokkum \& Franx 1996).
Similarly, for a single galaxy the offset from the Coma FP can be expressed
as an offset in $M/L$ ratio.
Offsets for individual galaxies in clusters at
$0.02\leq z\leq 1.27$ are shown in Fig.\ \ref{res.plot}(a).
The offsets in $\ln M/L_B$
for galaxies \gala\ and \galb\ are $-1.37 \pm 0.15$
and $-1.27 \pm 0.14$
respectively, remarkably consistent
with an extrapolation of results obtained at $0.02\leq z \leq 0.83$.
The offset of \galc\ is much larger at $\Delta \ln M/L_B = -2.58
\pm 0.17$, corresponding to a factor $\approx 13$.

In Fig.\ \ref{ml.plot} we investigate the
$M/L$ ratios of galaxies in \clusa\ in more detail by showing the
projection of the FP on the $M/L$ ratio vs.\ mass plane.
These quantities are defined in a straightforward way:
\begin{equation}
\log M \equiv 2 \log \sigma + \log r_e + C_1
\end{equation}
and
\begin{equation}
\log M/L_B \equiv 2 \log \sigma - \log I_e - \log r_e + C_2.
\end{equation}
With $\log I_e \equiv -0.4(\mu_{e,B} - 27.0)$, $r_e$ in kpc,
$C_1=6.07$ and $C_2=-1.29$, mass and $M/L$ ratio are expressed
in solar units (see J\o{}rgensen et al.\ 1996).

The mass of galaxy \galc\ is
a factor $\sim 10$ lower than that of \gala\ and \galb, and 
similar to the lowest mass galaxies in the ~{J\o{}rgensen} {et~al.} (1996)
Coma sample. We conclude
that its inclusion in our small sample is probably
due to its high surface brightness and low $M/L$ ratio, and
it is unlikely (although not impossible)
that this object is representative
for galaxies in the mass range $10 < \log M < 11$.

In Fig.\ \ref{res.plot} (b) we show
the evolution of the mean $M/L$ ratio of high mass galaxies,
by selecting objects with $M>10^{11}\,M_{\odot}$
in all clusters (solid symbols
in Fig.\ \ref{res.plot}a).
Note that this is not a true mass selected sample,
since we may miss ``underluminous''
galaxies with very high $M/L$ ratios.
The evolution is remarkably uniform and linear over the entire redshift
range $0.02 \leq z \leq 1.27$. A linear fit gives
$\ln M/L_B \propto (-1.06 \pm 0.09) z$, with the error dominated
by systematic uncertainties. The rms scatter around the best
fitting relation is only 0.03, significantly smaller than expected
from the uncertainties in the individual datapoints. This may partly
be caused by the cancellation of systematic errors, but can also be
the result of small number statistics. This is almost certainly true
for \clusa\ itself, because the random errors in the velocity
dispersions alone already introduce an uncertainty of $\sim 13$\,\%
in the average offset of \gala\ and \galb.

\subsection{Implications for Luminosity Weighted Ages}

The evolution of the mean $M/L$ ratio depends on the slope of the
Initial Mass Function (IMF), the metallicity, cosmological parameters,
and the luminosity weighted age of the stellar population ~(see, e.g.,
{Tinsley} \& {Gunn} 1976; {Worthey} 1994).  Furthermore, the observed
evolution needs to be corrected for the effects of morphological
evolution ~({van Dokkum} \& {Franx} 2001). The cited studies
provide a comprehensive discussion of these issues, beyond the scope
of the present paper. Here we only consider two models: a simple model
with a fixed IMF, varying age, and no morphological evolution, and the best
fitting complex model from ~{van Dokkum} \& {Franx} (2001).

\subsubsection{Simple Models}

In Fig.\ \ref{res.plot}(b) the dotted lines show predictions of simple
models with a ~{Salpeter} (1955) IMF, Solar metallicity, and a range
of formation redshifts. The models are of the form $L \propto
(t-t_{\rm form})^{\kappa}$, with $\kappa$ determined from stellar
population synthesis models ~(see {van Dokkum} {et~al.} 1998, and
references therein).  The formal best fitting formation redshift is
$z_* = 2.6^{+0.9}_{-0.4}$, corresponding to a luminosity weighted age
of $\sim 2$\,Gyr at the epoch of observation.  Note that these results do not
apply to the low mass galaxy \galc, which has a much lower luminosity
weighted age of $\sim 0.5$\,Gyr. This age is qualitatively consistent
with the presence of strong Balmer absorption lines in its spectrum.

The new data reinforce earlier studies at lower redshift.
Interestingly the formal error in the formation redshift of the stars
in massive galaxies suggests an upper limit of $z_*=3.5$ ($1\sigma$),
whereas previous studies could only provide lower limits. However, as
discussed in van Dokkum et al.\ (1998) the inferred formation
redshifts are strongly
dependent on the assumed cosmology and IMF.

\subsubsection{Complex Models}

The models discussed so far implicitly assume that
early-type galaxies in high redshift clusters are typical progenitors
of present-day early-type galaxies. However, there is
good evidence that the set of early-type galaxies evolves with time.
Dressler et al.\ (1997) and others (e.g., van Dokkum et al.\
2001b; Lubin, Oke, \& Postman 2002)
have found a gradual decrease with redshift of
the fraction of early-type
galaxies in clusters, from $\sim 80$\,\% in low redshift clusters
to $\sim 45$\,\% in clusters at $z\sim 1$, suggesting that
$\sim 50$\,\% of early-type galaxies was transformed from other
galaxy types in the last half of the age of the Universe.

A consequence of such morphological evolution is that the sample
of early-type galaxies at $z=1.3$ is only a subset of the full sample
of progenitors of present-day early-type galaxies. Furthermore,
the subset is biased, consisting of the oldest progenitors.
As a result the observed evolution of early-type galaxies underestimates
the true evolution that would be measured if all progenitors had been
accounted for.
van Dokkum \& Franx (2001) developed analytical
models which take this ``progenitor bias'' into account.
They can be used to correct the ages of early-type galaxies
for the effects of morphological evolution.
As shown in van Dokkum \& Franx (2001) a model with progenitor
bias can provide excellent simultaneous
fits to the observed evolution of the mean
$M/L$ ratio, the early-type
galaxy fraction, and the scatter in the color-magnitude relation
at $0.02\leq z \leq 0.83$. 
Their best fitting model is shown by the solid line in Fig.\
\ref{res.plot}(b). In this model
the mean luminosity weighted formation redshift
of the stars in all present-day early-type galaxies $\langle z_* \rangle
= 2.0^{+0.3}_{-0.2}$, whereas the formation
redshift of the stars in the subset of
early-type galaxies that were already assembled
at $z= 1.27$
is higher, at $\langle z_* \rangle \approx 2.5$.
As can be seen in Fig.\ \ref{res.plot}(b) the fit is equally good
as simple models with $z_* \approx 3$.

In summary, the $M/L$ ratios of the two galaxies with masses
$>10^{11}\,M_{\odot}$ are well fitted by extrapolations of models
fitted to data at $z<1$.  Good fits are obtained for simple models
that ignore selection effects and also for self-consistent models which
incorporate morphological evolution.


\section{Correcting for Luminosity Evolution at $z\approx 1.3$}

The observed evolution of the $M/L$ ratio can be used
directly to correct the luminosities of distant galaxies for
passive evolution, and convert luminosity to mass.
Ultimately measurements of $M/L$ ratios
at high redshift should remove the need for stellar population synthesis
models to interpret the evolution of the luminosity function,
and provide a direct measurement of the evolution of
the galaxy mass function.
The logarithmic
correction we derive from massive cluster galaxies
at $0.02\leq z \leq 1.27$ corresponds to
a brightening of $1.50 \pm 0.13$ magnitudes at $z=1.3$.
In absolute terms, a typical early-type galaxy with
$M/L_B = 5.9\,h_{50}$ in units of $(M/L_B)_{\sun}$ in the local Universe
~({van der Marel} 1991) has
$M/L_B = 1.5 \pm 0.2\,h_{50}$ at $z=1.3$.

Massive early-type galaxies are expected to be
a minor contributor to
$K$-selected samples of high redshift galaxies
~(e.g., {Cimatti} {et~al.} 2002), but they may form a large
fraction of the population of Extremely Red Objects. In Fig.\ \ref{col.plot}
the distance from the Coma Fundamental Plane, expressed in $R_z$
magnitudes, is plotted against $I-H$ color.
Interestingly, despite the low $M/L$ ratio of \galc\
all three galaxies have $I-H>3$ (and also $R-K>5$)
and fall in the class of Extremely Red Objects. As discussed
in \S\ 4.3 simple models
with $z_{\rm form} \approx 3$ can be used to correct
galaxies \gala\ and \galb\ for luminosity evolution,
but such models
severely underpredict the required correction for \galc.
Since EROs are usually not selected by mass but on the basis of
$K$-band luminosity, ERO samples cannot be pruned by applying a mass
cut (as we did in \S\ 4.2 for galaxies with measured
kinematics). Therefore, we conclude from our small sample that it is
hazardous to use simple ``Passive Luminosity Evolution'' (PLE)
models ~(e.g., {Pozzetti} {et~al.} 1996) to
correct EROs for luminosity evolution. Our results show that the
correction can range from $1-2.5$ magnitudes in the $H$-band at
$z\approx 1.3$, even in a restricted sample of luminous EROs with
early-type morphology. We note that it may be possible to improve the
reliability of the correction by including a color term (for galaxies
of known redshift); as demonstrated in Fig.\ \ref{col.plot} the
observed correlation between $\Delta H$ and $I-H$ color is similar to
the expected correlation from stellar population synthesis models.

\vbox{
\begin{center}
\leavevmode
\hbox{%
\epsfxsize=8.4cm
\epsffile{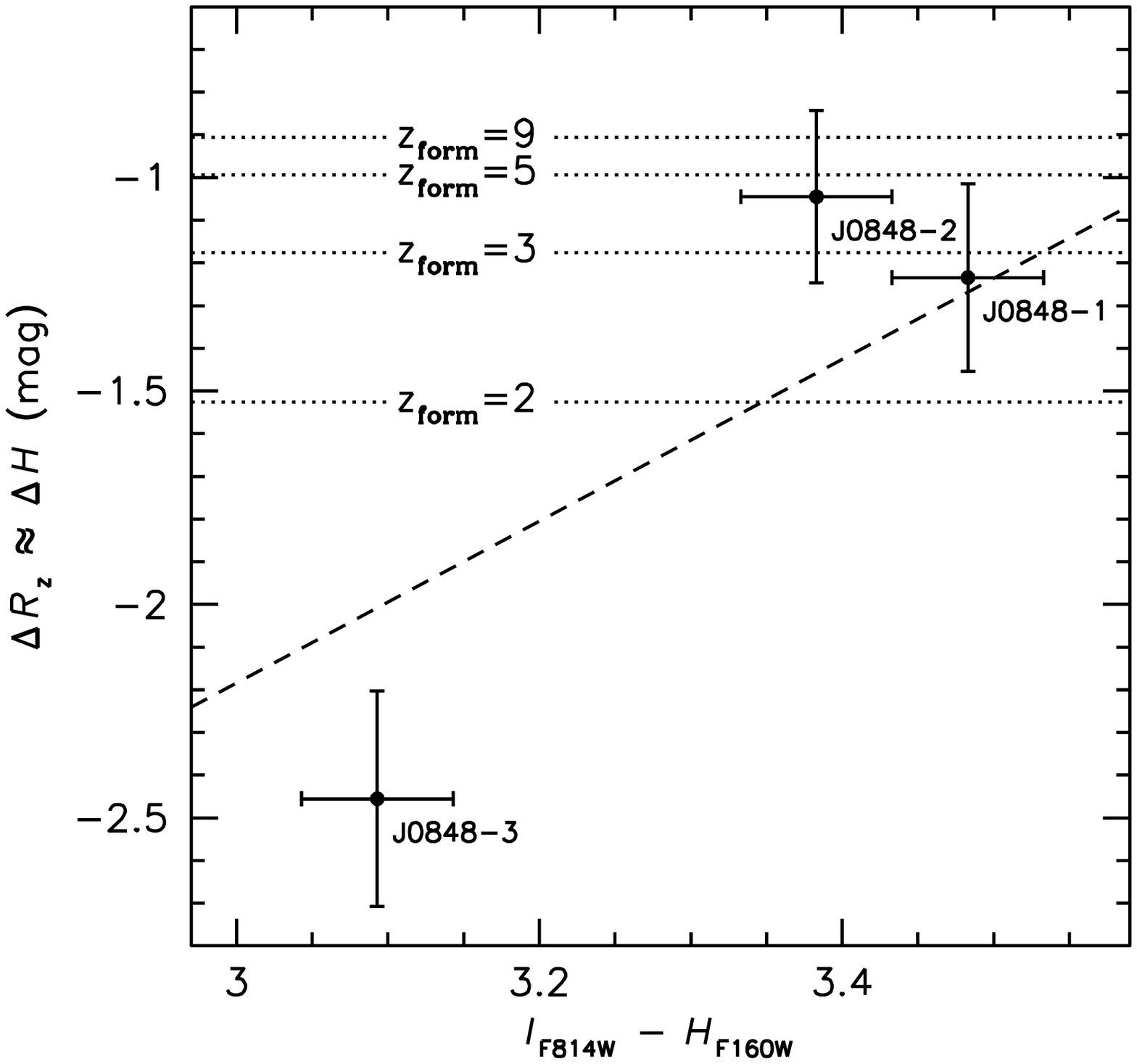}}
\figcaption{\small
Correction for luminosity evolution
in rest-frame $R$ (approximately
observed $H$) for
the three early-type galaxies at $z=1.27$. All three galaxies
have $I-H>3$, and fall in the class of ``Extremely Red Objects''.
Predicted corrections
for different formation redshifts of the stars
are indicated by dotted lines. The dashed line indicates the expected
relation between rest-frame $U-R$ color and $R$ magnitude for
a Salpeter (1955) IMF, solar metallicity, and varying age. Simple
``PLE'' models
with $z_{\rm form} \approx 3$ accurately predict the observed luminosity
evolution for two galaxies, but underpredict the luminosity
evolution of \galc\ by more than a magnitude.
\label{col.plot}}
\end{center}}

\section{Conclusions}

We have presented results on the Fundamental Plane of early-type galaxies
in the cluster \clusa\ at $z=1.27$. Although
our sample of three galaxies cannot be used
for a determination of the tilt
and scatter in the FP, we can
interpret the $M/L$ ratios of individual galaxies by comparing them
to the prediction of the FP in local clusters. 
The $M/L$ ratios of the
two most massive galaxies, when combined with those of massive galaxies
in clusters at $0.02 \leq z \leq 0.83$, are consistent with
a very regular evolution of the mean $M/L$ ratio of the form
$\ln M/L_B \propto (-1.06\pm 0.09) z$.
The third galaxy is a factor $\sim 10$ less massive than the other two,
and has a much lower $M/L$ ratio. Our data are insufficient to
determine whether the $M/L$ ratio of this galaxy is typical for its
mass. It is clear, however, that it is due to the presence of
a young stellar population:
the galaxy has an ``E+A'' spectrum, and its relatively blue
color is consistent with its low $M/L$ ratio.
Unfortunately it will be difficult to
obtain much larger samples in a single cluster at $z\sim 1.3$,
although it may be possible to combine results from
different clusters to constrain the  evolution of the FP better.

The overall conclusion from our work on \clusa\ is that by $z\sim 1.3$
we are approaching the epoch of formation of massive cluster galaxies.
In ~{van Dokkum} {et~al.} (2001b) we have shown that the most luminous galaxies
show signs of interactions, with the second brightest cluster galaxy
(a merger of three red $\sim L_*$ galaxies) the most spectacular
example. Furthermore, we presented evidence that the slope of the
color-magnitude relation may be flatter than in low redshift clusters.
The present paper shows that the spectra of three of
the most luminous early-type galaxies indicate the presence of
a young stellar population in one galaxy, and possibly residual ongoing star
formation in the other two. Furthermore, their $M/L$ ratios indicate
luminosity weighted ages of only $0.5$ -- $2$\,Gyr at the epoch of observation.

As discussed in \S\ 5 the measurements presented here are relevant in
a broader context than the evolution of galaxies in clusters, as they
can be used to correct the luminosities of distant red galaxies for
passive evolution of their stellar populations. Ultimately measurements
of the Fundamental Plane and Tully-Fisher relation at high redshift,
when combined with measurements of the luminosity function, should
lead to an empirical
determination of the evolution of the mass function. The sparse observations
that are available now are already valuable, since they can be used
to calibrate stellar population synthesis models and provide realistic
errors estimates. Apart from collecting larger samples in high redshift
clusters, an obvious next step is to obtain similar data on field galaxies
to determine whether there is an environmental dependence.

\acknowledgements{
We thank Mark Dickinson and Chris Hanley for the NICMOS reductions,
and Brad Holden for help with the PSFs. The detailed comments of the
anonymous referee improved the text.
P.\ G.\ v.\ D.\ acknowledges support by NASA through Hubble Fellowship
grant HF-01126.01-99A
awarded by the Space Telescope Science Institute, which is
operated by the Association of Universities for Research in
Astronomy, Inc., for NASA under contract NAS 5-26555, and through
the SIRTF Fellowship Program, administered by the
California Institute of Technology.
S.\ A.\ S.\ is supported by the Institute of Geophysics and Planetary Physics
(operated under the auspices of the US Department of Energy by the
UC Lawrence Livermore National Laboratory under
contract W-7405-Eng-48), and by NASA/LTSA
grant NAG5-8430. The authors wish to extend
special thanks to those of Hawaiian ancestry on whose sacred mountain
we are privileged to be guests. Without their generous hospitality,
many of the observations presented herein would not have been possible.
}

\end{document}